\documentclass{article}
\usepackage{amssymb}

\newtheorem{theorem}{Theorem}
\newtheorem{acknowledgement}[theorem]{Acknowledgement}

\begin{document}

\title{Time crystals: can diamagnetic currents drive a charge density wave
into rotation?}
\author{Philippe Nozi\` eres \\
Institut Laue Langevin, BP 156, F-38042 GRENOBLE Cedex9}
\maketitle

\begin{abstract}
It has been argued recently that an inhomogeneous system could rotate
spontaneously in its ground state - hence a 'time crystal' which is periodic
in time. In his note we present a very simple example: a superfluid ring
threaded by a magnetic field which develops a charge density wave (CDW). A
naive calculation shows that diamagnetic currents cannot drive rotation of
the CDW, with a clear picture of the cancellation mechanism.
\end{abstract}

In a recent letter \cite{Wilczek} Frank Wilczek introduced a revolutionary
concept, 'time crystals' which in their ground state would be time
dependent. A simple example is a ring threaded by a magnetic field that
breaks time reversal invariance. If a charge density wave (CDW) appears,
will diamagnetic currents put it in rotation? Such a challenging proposal
raised a vivid controversy \cite{Bruno1}, especially with my colleague
Patrick Bruno who recently provided a proof that it is impossible \cite%
{Bruno2}. In order to clarify the underlying physics, I study here a very
simple model in which elementary calculations can be done explicitly from
beginning to end.

Consider a circular ring with radius $R$, perimeter $L=2\pi R$, threaded by
a magnetic flux $\Phi .$The vector potential along the ring is $A=\Phi /L$.
As a model \ take a Bose condensate of charge $q$ particles (for Cooper
pairs $q=2e)$ with a density $n_{0}=N_{0}/L$. The superfluid phase is $S$,
the current density on the ring and the energy are 
\[
J=\frac{n_{0}}{m}\left[ \hbar \textrm{\,grad\,}S-\frac{qA}{c}\right] \;\;,\;E_{0}=%
\frac{N_{0}}{2m}\left[ \hbar \textrm{\,grad\,}S-\frac{qA}{c}\right] ^{2} 
\]%
The circulation of $\textrm{\,grad\,}S$ is quantified, equal to $2\pi \nu $ where $%
\nu $ is an integer. Linear momentum per particle is $p=\hbar \textrm{\,grad\,}S$,
without the gauge term. Angular momentum per particle is 
\[
L_{z}=R\hbar \textrm{\,grad\,}S=R\hbar \frac{2\pi \nu }{2\pi R}=\nu \hslash 
\]%
We recover usual quantization of angular momentum.

Consider first a weak magnetic field whose flux is smaller than half a
quantum: the ground state corresponds to $\nu =0$, with a diamagnetic
current $J=-n_{0}qA/mc.$ Assume now that the ring presents a spontaneous
charge density wave with a density modulation $n_{1}$. We do not need a
periodic $n_{1}$, we only request a modulation, $\oint n_{1}dx=0$.The
current equation becomes 
\[
J=\left( \frac{n_{0}+n_{1}}{m}\right) \left[ \hbar \textrm{\,grad\,}S-\frac{qA}{c}%
\right] 
\]%
$J$ is conserved and $\textrm{\,grad\,}S$ cannot vanish. We must solve the
equation 
\[
\hbar \textrm{\,grad\,}S=\frac{qA}{c}+\frac{mJ}{n_{0}+n_{1}} 
\]%
with the two conditions $\oint \textrm{\,grad\,}Sdx=0$ et $\oint n_{1}dx=0$
which will fix the modulation $\textrm{\,grad\,}S$ and the unknown current $J.$
Integrating that equation over the ring we find 
\[
\frac{qA}{c}+\frac{mJ}{n_{0}}\overline{\left( \frac{1}{n}\right) }=0 
\]%
Note that regions of small $n$ severely reduce the current $J$, a feature
pointed out long ago by Tony Leggett \cite{Leggett}. Here $n_{1}$ is small
and in lowest order we find 
\[
J=-\frac{n_{0}qA}{mc}\left[ 1-\frac{\overline{n_{1}^{2}}}{n_{0}^{2}}\right] 
\]%
The charge density wave reduces the diamagnetic current by an amount of
order $\overline{n_{1}^{2}}$. The ground \ state energy may be written as%
\[
E_{0}=\frac{1}{2}\oint J\left[ \hbar \textrm{\,grad\,}S-\frac{qA}{c}\right] =-%
\frac{eAJ}{2c}=\frac{N_{0}q^{2}A^{2}}{2mc^{2}}\left[ 1-\frac{\overline{%
n_{1}^{2}}}{n_{0}^{2}}\right] 
\]%
Note that the charge density wave \textit{reduces} the energy, a feature
that should eventually be added to the usual Landau picture for the
appearance of $n_{1}$%
\[
E_{CDW}=-\alpha n_{1}^{2}+\beta n_{1}^{4}\;\;\longrightarrow \;\;n_{1}^{2}=%
\frac{\alpha }{2\beta } 
\]
Diamagnetism \textit{enhances} the charge density instability.

Finally the angular momentum $L_{z}$, zero for a diamagnetic current in a
perfect ring,does not vanish any more when the density wave appears: 
\[
L_{z}=\oint \left( n_{0}+n_{1}\right) R\hbar \textrm{\,grad\,}S=\oint R\left[ \left(
n_{0}+n_{1}\right) \frac{qA}{c}+mJ\right] =N_{0}R\frac{qA}{c}\frac{\overline{%
n_{1}^{2}}}{n_{0}^{2}} 
\]%
Hence a crucial question: could such an angular momentum induce a
spontaneous rotation of the charge density wave? Standard wisdom says that
rotating the frame at angular velocity $\Omega $ adds a term $L_{z}\Omega $
to the energy: if true, rotation of the charge density wave is unavoidable
since all other terms in the energy are quadratic in $\Omega $. A wise
pedestrian approach is to stay in the laboratory frame when calculating
energy, noting however that in that frame the current is no longer
conserved. It is conserved in the rotating frame of the charge density wave:
we start from that statement and we bring everything back to the laboratory
frame.

Let $J^{\prime }$ be the constant current in the rotating frame. The
previous calculation relates it to a vector potential $A^{\prime }$ unknown
as of now 
\[
J^{\prime }=\left( \frac{n_{0}+n_{1}}{m}\right) \left[ \hbar \textrm{\,grad\,}S-%
\frac{qA^{\prime }}{c}\right] \;\Longrightarrow \;\hbar \textrm{\,grad\,}S=\frac{%
qA^{\prime }}{c}+\frac{mJ^{\prime }}{n_{0}+n_{1}} 
\]%
The current $J$ in the laboratory frame is 
\[
J=J^{\prime }+\left( n_{0}+n_{1}\right) \Omega R=\left( \frac{n_{0}+n_{1}}{m}%
\right) \left[ \hbar \textrm{\,grad\,}S-\frac{qA}{c}\right] 
\]%
where we have set $A^{\prime }=A+mc\Omega R/q$ :\ $J^{\prime }$ is related
to $A^{\prime }$ exactly as $J$ was to $A$ before. Rotating the charge
density wave is tantamount to replacing the vector potential $A$ by $%
A^{\prime }$ in the rotating frame$.$From there on the calculation \ unfolds
as before. The energy in the laboratory frame may be written as%
\begin{eqnarray*}
E_{0} &=&\oint \frac{n_{0}+n_{1}}{2m}\left[ \hbar \textrm{\,grad\,}S-\frac{qA}{c}%
\right] ^{2} \\
&=&\oint \frac{n_{0}+n_{1}}{2m}\left[ \left( \hbar \textrm{\,grad\,}S-\frac{%
qA^{\prime }}{c}\right) \left( \hbar \textrm{\,grad\,}S-\frac{qA^{\prime \prime }}{%
c}\right) +m^{2}\Omega ^{2}R^{2}\right] \\
&=&\frac{1}{2}\oint \left[ J^{\prime }\left( \hbar \textrm{\,grad\,}S-\frac{%
qA^{\prime \prime }}{c}\right) +n_{0}m\Omega ^{2}R^{2}\right]
\end{eqnarray*}%
where we have set $A^{\prime \prime }=A-mc\Omega R/q.$The constant current $%
J^{\prime }$ is 
\[
J^{\prime }=-\frac{n_{0}qA^{\prime }}{mc}\left[ 1-\frac{\overline{n_{1}^{2}}%
}{n_{0}^{2}}\right] 
\]
We find the energy%
\[
E_{0}=\frac{N_{0}}{2m}\oint \left[ \frac{q^{2}A^{\prime }A^{\prime \prime }%
}{c^{2}}\left( 1-\frac{\overline{n_{1}^{2}}}{n_{0}^{2}}\right) +m^{2}\Omega
^{2}R^{2}\right] 
\]%
The correction due to rotation is of order $\Omega ^{2}$. There is no linear
tem that could generate spontaneous rotation. Note that this second order
term vanishes if \ $n_{1}=0$: rotating something which \ does not exist
cannot cost any energy! In contrast rotating the density wave costs an
energy 
\[
E_{1}=\frac{mN_{0}}{2}\Omega ^{2}R^{2}\frac{\overline{n_{1}^{2}}}{n_{0}^{2}} 
\]%
The conclusion of this naive model is clear : \textit{a charge density wave
is not driven to rotation by diamagnetic current in the ground state }$\nu
=0 $

Generalization to an excited state $k=\nu /2\pi R$ is straightforward. We
still have%
\[
\hbar \textrm{\,grad\,}S=\frac{qA}{c}+\frac{mJ}{n_{0}+n_{1}} 
\]%
whose circulation is%
\[
\frac{\hbar \nu }{R}=\frac{qA}{c}+\frac{mJ}{n_{0}}\left[ 1+\frac{\overline{%
n_{1}^{2}}}{n_{0}^{2}}\right] 
\]%
The constant current $J$ and the energy become%
\[
J=\frac{n_{0}}{m}\left[ 1-\frac{\overline{n_{1}^{2}}}{n_{0}^{2}}\right] %
\left[ \frac{\hbar \nu }{R}-\frac{qA}{c}\right] \;,\;E=\frac{1}{2}\oint J%
\left[ \hbar \textrm{\,grad\,}S-\frac{qA}{c}\right] 
\]%
Since $J$ is constant only the circulation of $\textrm{\,grad\,}S$ matters, hence%
\[
E=\frac{N_{0}}{2m}\left[ \frac{\hbar \nu }{R}-\frac{qA}{c}\right] ^{2}\left[
1-\frac{\overline{n_{1}^{2}}}{n_{0}^{2}}\right] 
\]%
The only difference is the replacement of $eA/c$ by $\left( eA/c-\hbar \nu
/R\right) .$From \ there on \ the calculation is unchanged

\textit{Our final conclusion is clear, but limited: a charge density wave is
not driven to rotation in a quantum coherent state, for instance by a
diamagnetic current induced by a magnetic field, or by a persistent current
in a coherent, phase locked, superfluid state. }This is consistent with the
Ehrenfest theorem which states that the expectation value of the time
derivative of any observable $A$ is identically zero in any eigenstate $%
\left\vert \psi _{n}\right\rangle $of the hamiltonian:%
\[
\left\langle \frac{dA}{dt}\right\rangle =i\left\langle \psi _{n}\right\vert
AH-HA\left\vert \psi _{n}\right\rangle =0 
\]%
Any local motion of the charge density wave creates a local time dependence
which is precluded. The ring is a finite system and CDW motion is a local
issue.Such a conclusion holds for the ground state as well as for thermal
equilibrium where the density matrix is diagonal in the $\left\vert \psi
_{n}\right\rangle $ basis.This is no longer true if a current is forced in
the ring, breaking thermal equilibrium. Then CDW dragging becomes possible.

\begin{acknowledgement}
I wish to thank my colleagues Patrick Bruno and Andres Cano who introduced
me to the challenge of time crystals. Numerous discussions with them were
crucial in my search for simplicity. The reader is referred to the
forthcoming paper of Patrick Bruno \cite{Bruno2} which offers a much more
general proof, necessarily more elaborate.\ His conclusions are fully
consistent with my simple picture.
\end{acknowledgement}

\end{document}